%% file: GET-SI96.TEX
\documentclass[12pt]{iopart}

\usepackage{iopams}

\usepackage{graphicx}

\begin{document}

\title[]{Application of elastic mid-IR light scattering for
inspection of internal gettering operations
}

\author{O.~V.~Astafiev\footnote{E-mail: ASTF@KAPELLA.GPI.RU.}, A.~N.~Buzynin,
V.~P.~Kalinushkin\footnote{E-mail:
VKALIN@KAPELLA.GPI.RU.},
V.~A.~Yuryev\footnote{E-mail: VYURYEV@KAPELLA.GPI.RU.}}

\address{General Physics Institute of the Russian Academy of Sciences,
38, Vavilov Street, Moscow, GSP--1, 117942, Russia}

\begin{abstract}
A  method  of  low-angle   mid-IR  light scattering
is  shown  to  be   applicable   for   the contactless   and
non-destructive  inspection  of  the internal  gettering
process   in  CZ Si crystals.
A classification of scattering inhomogeneities in
initial
crystals and ones subjected to the getting process  is
presented.
\end{abstract}

\section{Introduction}
 Recently, the internal  gettering  process has become  one  of  the  main
operations for manufacturing of semiconductor  devices of CZ Si.
However, methods for the direct  inspection  of
 the internal gettering efficiency and stability   have  been  practically
absent  thus  far.  The  purpose  of  this   paper is to
present such a method developed on the basis of law-angle
mid-IR-light  scattering  technique  (LALS)  \cite{1},  which  has   been
successfully applied thus far for the investigation  of  large-scale
electrically active defect accumulations (LSDAs) in semiconductor
crystals (see e.g. \cite{2} and references therein).

A method of LALS was applied at the first time to the investigation
of the influence of both the internal and abrasive gettering processes
on large-scale impurity accumulations (LSIAs\,\footnote{LSIAs are a type
of LSDAs
which contain mainly impurities rather than intrinsic defects.})  in
crystals of the industrial CZ Si:B in Ref.\,\cite{3}. The conclusions
were made there that (i) the abrasive gettering process
resulted in a considerable decrease of the impurity concentration
in LSIAs and (ii) new defects arose in the crystal bulk as
a result of the internal
gettering process which became a predominating type of defects.
Unfortunately, parameters of these new defects were not determined
in Ref.\,\cite{3}.

The current work presents an application of LALS with the non-equilibrium
carrier photoexcitation \cite{1,4} to the studies of the internal
gettering process in addition to the conventional LALS measurements.

The LALS temperature dependances are also presented and the activation
energies of the centers constituting the LSIAs are estimated in this work.

\section{Experimental}
A continuous  10.6-$\mu$m emission of a
CO$_2$-laser was  used  as  the  source  of  the probe  radiation in LALS.
All the details of this technique are described in Refs.\,\cite{1,2}.
We would like to remind only that such parameters of LSDAs as
their effective sizes and the product of the LSDA concentration by
\begin{figure}[t] 
\begin{center}
\includegraphics[scale=1.8]{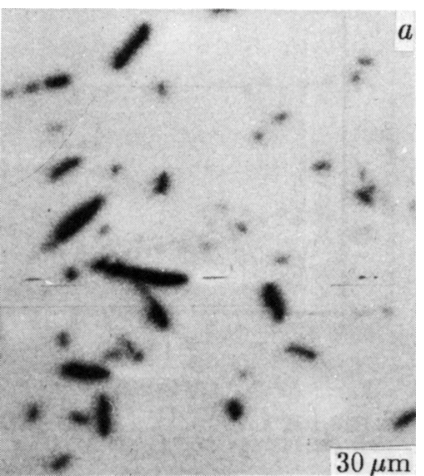}
\includegraphics[scale=1.8]{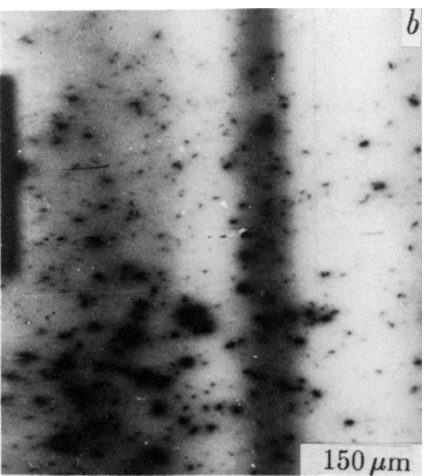}
\end{center}
\caption{Typical EBIC pictures of impurity accumulations in the
as-grown sample $(a)$ and after the internal gettering process
$(b)$.}\label{f1}
\end{figure}
the square of the deviation of the free carrier concentration (or
the square of the dielectric constant variation) in LSDAs
$C\Delta n^{2}_{ac}$
(or $C\Delta \varepsilon_{ac}^{2}$) can be calculated from the
light-scattering diagrams.

The investigation of the influence of a sample temperature on its
light scattering enables the estimation of the thermal activation
energies ($\Delta E$) of impurities and defects composing the
LSDAs\,---\,$I_{sc}\sim\Delta n_{ac}^{2}$ \cite{5}.
During the low-temperature measurements,  the sample temperature
varied from 80 to 300K.

The  influence  of  non-equilibrium  carrier
photoexcitation on light  scattering  was  studied  as  well.  The
essence of the experiments consist in the following.  If  a  crystal contains
large-scale centers of recombination (e.g. precipitates and
their colonies, stacking faults, swirls, {\it etc.}), regions with decreased
concentration of non-equilibrium carriers
are  formed around these  centers during the process  of
the non-equilibrium carrier generation.  These
regions scatter the mid-IR light like usual non-uniformities \cite{1,4}
and when  pulse generation of carrier
is used,  the  scattered  light  pulses  are  observed in LALS.
Selecting this pulsed component, it is possible to register the
light scattering by recombination defects (RDs).  Then the usual
procedure of the light-scattering diagram measurement and treatment is
applied to estimate the dimensions of the depleted regions around
RDs \cite{1,4}.
In this work, the non-equilibrium carrier  was  generated
by 40-ns pulses of YAG:Nd$^{3+}$-laser  at the wavelength of 1.06\,$\mu$m,
frequency of 1\,kHz and  mean  power of 1\,W. The photoexcitation at this
wavelength pumps whole the crystal bulk practically uniformly, as the
absorption for this wavelength is not too high but sufficient to produce the
efficient enough electron-hole pair generation.
The scheme of  the  used  instrument is described in detail in
Refs.\,\cite{1,4}.

Electron beam induced current  (EBIC)
and selective etching were used to reveal the defects as well.
During  the  sample  preparation  for   EBIC,   a special
\begin{figure}[t]
\begin{center}
\includegraphics[scale=4]{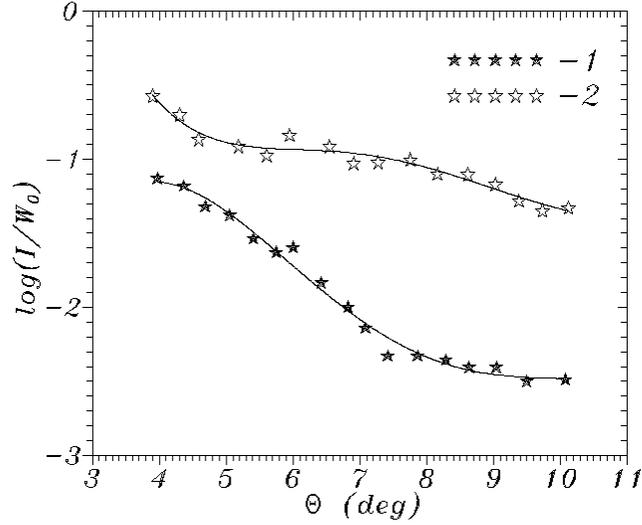}
\end{center}
\caption{Characteristic light-scattering diagrams for CZ~Si:B:
(1) as-grown, (2) after internal gettering.}\label{f2}
\end{figure}
technique was  used which included the plasma etching of the sample
surface in special regime before the Schottky barrier was created.
This  technique  considerably increases  EBIC  sensitivity to
RDs in bulk Si \cite{6}.

 About 40 wafers of dislocation-free Si  were  studied.
The crystals were produced by Czochralski method and doped with boron
(CZ Si:B) up to the specific resistivity  from  1  to  40~$\Omega$\,cm.
The  crystals  were
produced at three different establishments and subjected to the internal
gettering process at five different establishments.
In this paper, the  effect of  different  gettering  regimes on
defects is  summarized.
In the experiments on LALS, the two following schemes of experiment  were
used. In the first scheme, a preliminary study of  the as-grown  substrates,
which then were subjected to the internal gettering process and examined by
LALS, was  carried  out.
In the other one, the substrates were cut into several  parts.  Some
of them were subjected to the gettering process and the others
were used as reference samples.

Experiments on  EBIC and selective etching were carried out only in accordance with
the second scheme.

\section{Results}
\subsection{As-grown samples}
          The studied here initial wafers contained a
standard for CZ Si:B set of LSIAs \cite{7}. In these
samples, so-called  {\it cylindrical  defects}  (CDs)  with the
lengths from   15 to 40~$\mu$m   and   diameters  from 5 to 10~$\mu$m
 were    observed in the EBIC microphotographs
\begin{figure}[thb]
\begin{center}
\includegraphics[scale=4]{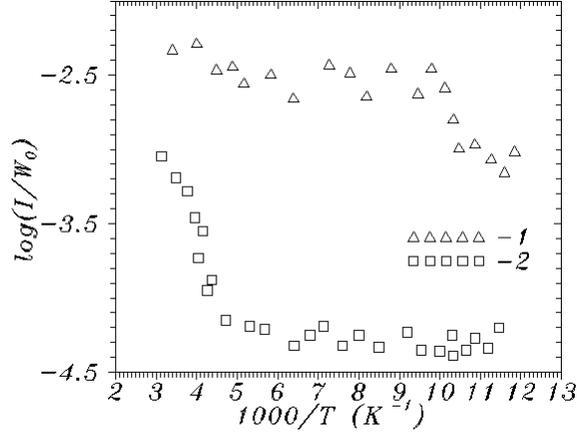}
\end{center}
\caption{Temperature dependances of the light scattering intensity
by the cylindrical (1) and spherical (2) defects in the
as-grown CZ~Si:B.}\label{init}
\end{figure}
(Fig.\,\ref{f1}\,$(a)$). The   concentration   of   CDs was
estimated as 10$^6$ to 10$^7$\,cm$^{-3}$.
In the scattering diagram (Fig.\,\ref{f2}, curve 1),
these defects correspond  to the sections at $\theta < 7^{\circ}$.
So-called {\it spherical defects} (SDs), the concentration  of
which in initial  samples  did  not  exceed  10$^5$\,cm$^{-3}$,  were  also
observed (Fig.\,\ref{f1}\,$(a)$).
The dimensions of these defects were from 5 to 15~$\mu$m.
These defects are manifested as the sections at $\theta >
7^{\circ}$ in the scattering diagram (Fig.\,\ref{f2}, curve~1).

The thermal activation energies of point centers controlling the light
scatter by the cylindrical and spherical defects evaluated from the
IR-light-scattering intensity temperature dependances (Fig.\,\ref{init})
ran from 40 to 60~meV and from 120 to 160~meV, respectively~\cite{7}.

 The depleted regions around RDs in initial  samples  mainly
had the dimensions less than 4\,$\mu$m and looked  as  a
``plateau'' in  the  diagrams (Fig.\,\ref{f3}, curve 1, $\theta > 5^{\circ}$).
Sometimes the SDs were also observed (Fig.\,\ref{f3}, curve 1,
$\theta < 5^{\circ}$).

\subsection{After internal gettering}
 The main result for the crystals subjected to the internal gettering
process are as follows:

1. As  a  result  of  the internal  gettering,   SDs  with the
dimensions from 10 to 30~$\mu$m (Fig.\,\ref{f1}$(b)$;
Fig.\,\ref{f2}, curve  2)  became
\begin{figure}[thb]
 \begin{center}
\includegraphics[scale=4]{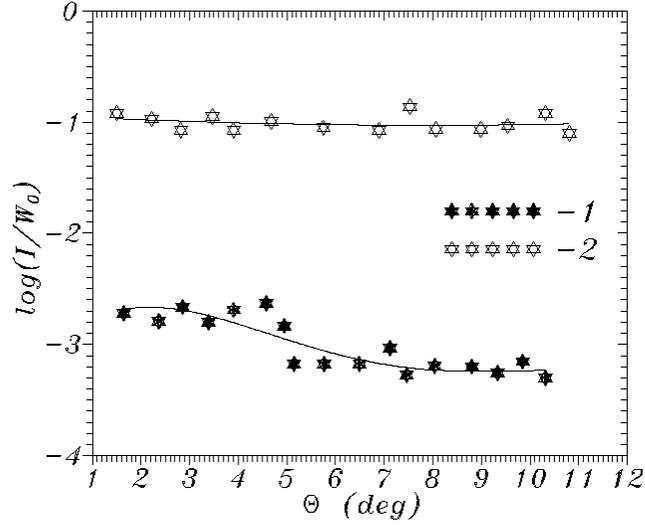}
\end{center}
\caption{Characteristic light-scattering diagrams for RDs in
CZ~Si:B measured with photoexcitation: (1) as-grown,
(2) after internal gettering.}\label{f3}
\end{figure}
the predominating type of defects  and  $I_{sc}$ for  these  defects
became by two orders  of  magnitude  greater  in comparison with that for the
initial material. $I_{sc}$ for CDs changed  rather  weakly  and  this
led to prevalence of scattering by SDs in the scattering diagrams of the
substrates subjected to the internal gettering.
We can conclude that the concentration of SDs  considerably increased
as a consequence of the internal gettering (Fig.\,\ref{f1}\,$(b)$).
The increase of $I_{sc}$ and the SD concentration  after the internal gettering
was  the  common
phenomenon for all the studied samples. This phenomenon  did  not
depend on the gettering regime. The increase of SD-related  $I_{sc}$
correlated with the appearance  of the gettering  defects  revealed
by selective etching. We did not obtain a
proportional dependance of SD-related $I_{sc}$ on epd, however .

2. It was found that RD-related $I_{sc}$ significantly increased after the
internal gettering (Fig.\,\ref{f3}, curve 2) and  a good
correlation  of  $I_{sc}$ with  epd  was
observed.

\begin{figure}[thb]
 \begin{center}
\includegraphics[scale=3.5]{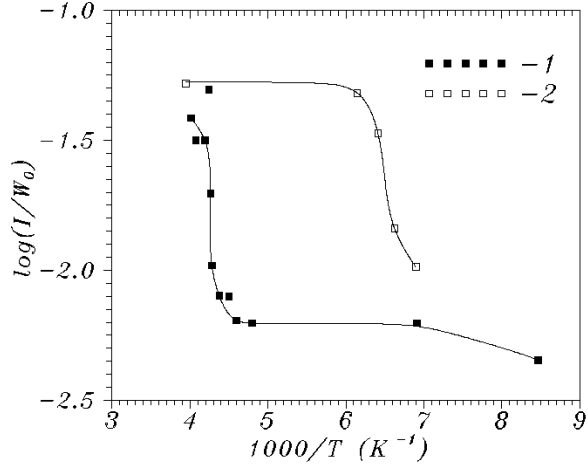}
\end{center}
\caption{Temperature dependances of the light-scattering intensity for
two different CZ Si:B samples after the internal gettering process.}\label{f4}
\end{figure}

3. The values of the activation energies  ($\Delta E$)  of the centers
predominating in SDs, perhaps,  are  defined  by  the  growth
conditions and the thermal prehistory of a sample. For  example
in Fig.\,\ref{f4}, the SD-related $I_{sc}$ temperature dependances are
shown for two
samples grown at different establishments and subjected to the internal
gettering.  It  is  seen  that these dependances as well
as  the values of $\Delta E$
are absolutely different for these samples, the estimates
gave the values of $\Delta E$ to be from 130 to 170~meV
for the sample 1 and  from 60 to 90~meV for the sample 2.
So, we can conclude that different point
centers constituted SDs in these samples after
the internal gettering.  Nowadays
we have not got enough data to state  which
of the following factors defines
the  SD composition: the initial pollution of the
materials by residual impurities, their thermal history
or the gettering process peculiarities.
We suppose the first two factors to be more essential.

\section{Discussion}
On the basis of the above results we assume  RDs  in the
substrates subjected to the internal gettering
to be some defects of structure  (most  likely, they are
precipitates and their colonies) which are formed
in the wafer bulk during the gettering defect formation
process,  and  are  the  gettering
defects themselves. As for SDs, we suppose them
to be the impurity atmospheres  around
these defects. These atmospheres are formed  when  impurities
flow to the gettering defects\,\footnote{It is
important to note that according to the model proposed SDs
consist at least of two components: the dissolved
impurities and the impurity precipitates. In the conventional LALS
experiments\,---\,without photoexcitation\,---\,only the
first component is observed. The second component
being recombination active is manifested in LALS
with the excess carrier photoexcitation.}. These conclusions are confirmed
by the correlation of RD-related $I_{sc}$ with epd and the  increase  of
SD-related $I_{sc}$ after the appearance of a great number of defects
revealed by selective etching.

 For any inhomogeneity, $I_{s c}$ is proportional to
$C\Delta \varepsilon _{ac}^{2}$
where $C$ is the
defect concentration and $\Delta \varepsilon _{ac}$  is  the
deviation of the dielectric constant inside them. In the case  of  RDs,
$\Delta \varepsilon _{ac}$ is determined  first of all  by  the
generated  excess carrier concentration. At equal levels of
photoexcitation (non-equilibrium carrier concentration),
RD-related light scattering intensity $I_{sc}$
is proportional to the RD concentration $C$ for the crystals with different
$C$. So $I_{sc}$ in this case is a straightforward measure of the
RD concentration in crystals.

In  the  case  of  SDs,
the situation is much more sophisticated. $\Delta \varepsilon _{ac}$
in them is controlled by many parameters:
the average concentration  of  impurities  in  SDs,
the ratio  of the dissolved  and   precipitated   impurity concentrations,
the compensation degree, {\it etc}. The gettering process may change both
$C$ and any of these parameters which will
result in the seeming violation of the $I_{sc}$ proportionality to
$C$ in experiments. Hence in the case of SDs, we are forced to  rely
only upon a qualitative correlation which  is  observed  in the
experiment\,\footnote{Now new techniques of scanning LALS (SLALS and OLALS)
have been developed
which enable the direct visualization of defects and accurate estimation of
their parameters \cite{8}. These techniques are very promising for
the internal gettering investigation \cite{9}.}.

On the basis of the above, we suggest the following
method for the examination of the gettering process in silicon
substrates.

The RD-related light scattering measurements (LALS with photoexcitation)
enable the inspection for the presence
of  the gettering  defects in the substrate
bulk.

The investigation of  SD-related light scattering  (conventional LALS
measurements) enables the
inspection the presence  and  parameters  of  the impurity
atmospheres around the gettering defects, i.e.  the efficiency and
stability of the gettering operation.

Note that the main advantage of this  technique
is  its applicability for  the  input  and
technological step inspection of substrates during the whole  technological
cycle.  It enables the examination of efficiency and
stability of  the gettering  process  after  any
high-temperature operations. The equipment may be easily  adapted
for the technological process; it allows one to carry out
the express (for 1 or 2 minutes) testing and mapping of  substrates
of practically any diameter.

The   studies of $I_{sc}$  temperature  dependances  enable the
analysis of  the impurity atmospheres  composition,  i.e.  they allows one
to  determine
what impurities are gathered by the gettering defects from
the denuded zone. In this case it is
reasonable to  speak  of  a random  inspection  or  laboratory
researches.

\section{Conclusion}
          In conclusion,  we would like to pay your attention  upon
an additional  very promising potentiality of the LALS-based
methods.

 It appears to be possible to make the nondestructive
input and technological step
inspection for the presence of RDs not only in a wafer bulk  but
directly  in  the  denuded  zone.  The  proposed  technique  is
analogous to the method of RDs revealing in a substrate bulk, but
instead of the ``bulk'' excitation of the electron-hole pairs it is proposed
to use
the ``surface'' photoexcitation (e.g. using short pulses of
the second  harmonic  of
 YAG:Nd$^{3+}$-laser,  $\lambda =0.53$\,$\mu$m).  In  this  case,
the excess carriers will penetrate
into the  depth
of 10 to 20-$\mu$m subsurface layer, and just  this  layer  will  be  analyzed
for  the
content  of  recombination-active  defects  of   structure   and
precipitates. The main difficulty of this approach
consists in rather small light scattering intensity $I_{sc}$  but
the preliminary  experiments  demonstrated  the  possibility  of
registration of light  scattering  from   layers  with the  RD
concentration down to 10$^4$--10$^5$\,cm$^{-3}$ \cite{10}.
This technique is undoubtedly very
attractive since it will allow one to directly determine  a  degree  of
purification of the denuded zone (including the minority
carrier lifetime measurements)
rather than study processes  developing
in the wafer bulk. For instance, it would enable the  inspection  of the
effect of precipitates ``germinating'' in the denuded  zone  during
the technological  cycle.  Note  also  that  this  technique  is
suitable not only for solving the problems of  gettering  but
also for the examination of any epitaxial and boundary
layers\,\footnote{The visualization of RDs in subsurface layers
is also possible now using the technique of the scanning
mid-IR-laser microscopy described in Refs.\,\cite{8,9}}.

\section*{Acknowledgments}

This work was partially made under the grant of the Russian
Foundation for Basic Researches No.~96-02-19540. The authors thank
RFBR for the support.
\\

\cleardoublepage

\input{SI_96GET.TEX}

\end{document}

%% file: SI_96GET.TEX
\oddsidemargin  0.25in
\evensidemargin 0.25in
\baselineskip 1em
\textwidth 160mm
\textheight 240mm
\pagestyle{empty}
\parskip=0ex
\parindent=4ex


\begin{center}
\Large
\bf

APPLICATION OF ELASTIC MID-IR-LIGHT SCATTERING
FOR INSPECTION OF INTERNAL GETTERING OPERATIONS\\[3ex]

O.~V.~Astafiev, A.~N.~Buzynin, V.~P.~Kalinushkin, and V.~A.~Yuryev \\[2ex]

\small
\sl
General Physics Institute of RAS, 38, Vavilov Street, Moscow, 117942,
GSP-1, Russia \\[1ex]
Tel./Fax: 7\,(095)\,135\,13\,30;\\ [0.5ex]
E-mail:
vkalin@ldpm.gpi.ru\\[3ex]

\end{center}
\normalsize
\rm

   In the present work, the possibilities of application of  elastic
scattering of CO$_2\,$-laser radiation for the express contactless incoming
and technological step inspection of the internal gettering efficiency
in CZ Si single crystals are considered on the basis of a  wide
experimental material.\\

The main conclusions are:
\begin{enumerate}
\item
The presence and stability of the gettering  precipitates  can
be checked by use of the elastic scatter of CO$_2\,$-laser light on  the
domains with decreased concentration of inequilibrium carriers  that
arise around these precipitates when the inequilibrium carriers  are
generated in the crystal bulk using Nd:YAG laser emission.$\sl {^1}$
\item
The presence of impurity atmospheres around these precipitates
can be registered using the standard technique of low-angle
mid-IR-light scattering (LALS),$\sl {^1}$ and the composition of the impurity
atmospheres can be found out measuring the LALS intensity dependencies
on the crystal temperature.$\sl {^2}$
\item
The level of clearing of near surface ``working'' layers of
substrates can be tested using the mid-IR-laser microscope$\sl {^3}$ in the
optical beam induced scattering mode, which is described in the
report cited in {\sl Ref.\,[4]}. The latter method enables to visualize
the  defects remained in the near-surface layers after the internal
gettering process.\\[2ex]
\end{enumerate}
\small
\begin{flushleft}
$\sl {^1}$V.~P.~Kalinushkin, {\em  Proc.\,Inst.\,Gen.\,Phys.\,Acad.\,Sci.\,USSR},
Vol.\,4,  New York, Nova, 1988, p.1.\\
$\sl {^2}$V.~P.~Kalinushkin, V.~A.~Yuryev,  D.~I.~Murin,  and M.~G.~Ploppa,
{\em Semicond. Sci. Technol.} {\bf 7} (1992) A\,255.\\
$\sl {^3}$O.V.Astafiev, V.P.Kalinushkin, and V.A.Yuryev, {\em Proc. SPIE}
{\bf 2332} (1994) 138.\\
$\sl {^4}$O.~V.~Astafiev, V.~P.~Kalinushkin, and V.~A.~Yuryev,
{\em Proc. DRIP--VI} (1996).\\
\end{flushleft}

%% file: GET-SI96.bbl
\begin{thebibliography}{10}
\bibitem{1}
V.~P.~Kalinushkin {\em Proc.~Inst.~Gen.~Phys.~Acad.~Sci. USSR}
vol.~4 {\em Laser Methods of Defect Investigations in Semiconductors and
Dielectrics} (1988) (New York: Nova)  1--79
\bibitem{2}
V.~P.~Kalinushkin, V.~A.~Yuryev and O.~V.~Astafiev {\em Proc.~1st Int. Conf.
on Materials for Microelectronics, Barcelona, 17--19 October 1994} in:
{\em Mater. Sci. Technol.} in press
\bibitem{3}
E.~N.~Gulidov, V.~P.~Kalinushkin, D.~I.~Murin \etal {\em Sov.
Phys.--\,Microelectronics} {\bf 14} (2) (1985) 130--133
\bibitem{4}
V.~P.~Kalinushkin, D.~I.~Murin, T.~M.~Murina \etal  {\em Sov.
Phys.--\,Microelectronics}  {\bf 15} (6) (1986) 523--527
\bibitem{5}
V.~P.~Kalinushkin, V.~A.~Yuryev, D.~A.~Murin\  and M.~G.~Ploppa~M~G  {\em Semicond.
Sci. Technol.}  {\bf 7} (1992) A255--A262\\
V.~P.~Kalinushkin, V.~A.~Yuryev\ and  D.~I.~Murin {\em Sov. Phys.--\,Semicond.}
{\bf 25} (5) (1991) 798--806\\
V.~V.~Voronkov, G.~I.~Voronkova, V.~P.~Kalinushkin \etal
{\em Sov. Phys.--\,Semicond.} {\bf 18} (5) (1984) 938--940\\
S.~E.~Zabolotskiy, V.~P.~Kalinushkin, D.~I.~Murin \etal
{\em Sov. Phys.--\,Semicond.} {\bf 21} (8) (1987) 1364--1368
\bibitem{6}
A.~N.~Buzynin, N.~A.~Butylkina, A.~E.~Lukyanov \etal
{\em Bul. Acad. Sci. USSR. Phys. Ser.} {\bf 52} (7) (1988) 1387--1390
\bibitem{7}
A.~N.~Buzynin, S.~E.~Zabolotskiy, V.~P.~Kalinushkin \etal
{\em Sov. Phys.--\,Semicond.} {\bf 24} (2) (1990) 264--270\\
O.~V.~Astafiev, A.~N.~Buzynin, A.~I.~Buvaltsev \etal
{\em Sov. Phys.--\,Semicond.} {\bf 28} (3) (1994) 407--415\\
V.~P.~Kalinushkin, A.~N.~Buzynin, V.~A.~Yuryev \etal
{\em Inst. Phys. Conf.
Ser.} {\bf 149} (1996) 219--224
\bibitem{8}
O.~V.~Astafiev, V.~P.~Kalinushkin\  and V.~A.~Yuryev
{\em Proc. Soc. Photo-Opt. Instrum. Eng.} {\bf 2332}
(1994) 138--145\\
O.~V.~Astafiev, V.~P.~Kalinushkin\  and V.~A.~Yuryev
{\em Inst. Phys. Conf. Ser.} {\bf 146} (1995) 775--780\\
O.~V.~Astafiev, V.~P.~Kalinushkin\  and V.~A.~Yuryev
{\em Mater. Sci. Eng.} B {\bf 34} (1995) 124--131\\
O.~V.~Astafiev, V.~P.~Kalinushkin\  and V.~A.~Yuryev
{\em Russian Microelectronics} {\bf 25} (1) (1996) 41--53\\
O.~V.~Astafiev, V.~P.~Kalinushkin\  and V.~A.~Yuryev
{\em Inst. Phys. Conf. Ser.} {\bf 149} (1996) 361--366
\bibitem{9}
O.~V.~Astafiev, V.~P.~Kalinushkin, V.~A.~Yuryev, A.~N.~Buzynin\
and N.~I.~Bletskan {\em Mater. Res.
Soc.  Symp. Proc. Vol.} {\bf 378} (1995) 615--620
\bibitem{10}
V.~P.~Kalinushkin, D.~I.~Murin, V.~A.~Yuryev, O.~V.~Astafiev\  and
A.~I.~Buvaltsev {\em Proc. Soc. Photo-Opt. Instrum. Eng.}\  {\bf 2332}
(1994) 146--153 \\
O.~V.~Astafiev, A.~I.~Buvaltsev, V.~P.~Kalinushkin, D.~I.~Murin\  and
V.~A~Yuryev {\em Phys. Chem. Mech. Surf.} (4) (1995) 79--83
\end{thebibliography}
